\documentstyle[12pt]{article}
\normalbaselines
\begin{document}

\begin{center}
{\LARGE {\bf Evaporating black holes and long range scaling}}\\

\vskip 1.5cm

{\large {\bf Hadi Salehi\footnote { Electronic address:
salehi@netware2.ipm.ac.ir}}}
\vskip 0.5 cm
 Department of Physics, Shahid Beheshti University, Evin,
Tehran 19834,  Iran.\\
\end{center}

 \vspace{2.5 cm}

\begin{abstract}

For an effective treatment of the evaporation process of a large
black hole the problem concerning the role played by the
fluctuations of the (vacuum) stress tensor close to the horizon
is addressed. We present arguments which establish a principal
relationship between the outwards fluctuations of the stress
tensor close to the horizon and quantities describing the onset of
the evaporation process. This suggest that the evaporation
process may be described by a fluctuation-dissipation theorem
relating the noise of the horizon to the black hole evaporation
rate.

\end{abstract}

\vspace{1.5 cm}
\section{Introduction}

One of the central questions in the theory of black hole
evaporation concerns the detailed understanding of the
characteristics scales involved in the treatment of the onset of
evaporation process. In the usual treatment the characteristic
scale of length for a black hole of mass $M$ is identified with
the Schwarzschild radius\footnote{We use units in which
$G=c=\hbar=1$.} $2M$ which for a large black hole is a macroscopic
length. Therefore, for a large black hole one may be inclined to
persist on the paradigm of an effective description requiring a
low energy treatment of black hole evaporation involving only the
characteristic energy scale $\sim\frac{1}{2M}$.

The point, however, is that due to the infinite gravitational
red-shift on the horizon, the long time (and long distance)
observations in the outside region of a black hole exhibit
correlations with the physical situations in a high energy regime
in the vicinity of the horizon where the fluctuations of the
(vacuum) stress tensor due to the high energy gravitational
effects can no longer be neglected [1][2][3]. This remark
demonstrates that a low energy description involving only the
characteristic scale $\frac{1}{2M}$ may not encompass the
essential features of black hole evaporation. In fact it seems
that high energy gravitational effects may profoundly affects the
correct form of the stress-tensor fluctuations near the horizon
in such a way that applicability of semiclassical methods for the
treatment of outside observations becomes questionable.

A promising idea towards clarifying this issue comes from the
application of the black hole complementarity principle [4] [5],
which emphasizes that the role played by the high energy effects
near the horizon depends essentially on whether we look at
physical states from the outside or the inside of the horizon.
For example, in the Hilbert space used in the description
of an outside static observer, the principle demands that the
high energy gravitational effects decouple themselves in form of a materialized
stretched horizon where the incoming information are transferred
into the outgoing thermal radiation. Thus as long as physical
states are strictly localized outside the (stretched) horizon one may assume
that the high energy gravitational
effects are suppressed in such a way that the correct
form of the stress-tensor fluctuations near the horizon can be
taken to posses only the appropriate low energy characteristics of a
small disturbance of black hole causal structure, so that
semiclassical methods may be applied to a good approximation.
The conditions of this effective Hilbert space outside the horizon
however can not be indicative of the observations made by an
infalling observer crossing the horizon because the high energy
effects never seem to decouple from the inside of the horizon so
that the Hilbert space becomes the wrong Hilbert space for the
fundamental low energy description of the physical state of an
infalling observer. To avoid any physical inconsistency one
requires that there is basically no way to combine the
description of the outside observations with the description of
observations made by infalling observers crossing the horizon,
the two descriptions are complementary descriptions. In this way
complementarity is claimed to reflect an important feature of
black hole evaporation.

For a detailed understanding of the principle of black hole
complementarity it is important to realize that the principle
essentially implies two things. Firstly it implies an assertion
about the impossibility of realizing the physical state of an
outside static observer and the physical state of an infalling
observer crossing the horizon in the same Hilbert space. In fact,
the two states are related to basically different sets of
boundary conditions inside the horizon. This is a statement about
the complementary properties of basically different Hilbert spaces
used by observers separated by mutually exclusive behavior of
their coordinates on the horizon.
The second implication is that for the effective description of
observations outside the horizon the principle requires a
systematic link between a small disturbance of black hole causal structure
and the evaporation process through the choice of a
physical low energy state. In order to establish such a link we
shall study a model in which the stress-tensor fluctuations near
the horizon, expressed in coordinates of an outside static
observer, are taken to posses the appropriate low energy
characteristics of random fluctuations arising from the noise
of gravitational effects inside the horizon (the horizon noise).
According to the principle of black hole
complementarity this effective ansatz, which is studied in this
paper, is correct as long as the physical state of an static
observer is strictly localized outside the horizon. Inside the
horizon the methods of this effective ansatz are generally felt
to break down, because the correct form of the physical state of
an static observer inside the horizon never seem to posses
the low energy characteristics required by that effective ansatz.

This effective ansatz is very useful for posing a general
question concerning the dissipative effects of quantized fields
in the presence of a black hole. We would expect, namely, that
the random fluctuations of the stress tensor near the horizon,
expressed in coordinates of an outside static observer, to posses
a dissipative character. This kind of behavior is suggested in a
very general way by fluctuation-dissipation theorems which
systematically link the random fluctuations of a system to a
systematic effect, namely the dissipative behavior of the same
system over long time intervals. In the present context the
dissipation is represented by the black hole evaporation process.
Therefore it is important to determine how the random fluctuations
of the stress tensor near the horizon can be related by the
conditions of a semiclassical theory to the evaporation rate. In
this paper we shall study this relationship using a
two-dimensional Schwarzschild black hole model. The significance
of such a lower dimensional model lies in the fact that it may be
considered as a model arising from the geometric optics
approximation of a physical spherically symmetric black hole
model. Such a restriction to geometric optics approximations and
to the corresponding lower dimensional methods simplifies
considerably the analysis, and it is generally believed that the
qualitative features of the evaporation process will not alter
too much by this restriction. The organization of the paper is as
follows: In the subsequent two chapters we present the model and
discuss heuristic arguments leading to  a long range scaling law
which controls the outwards stress-tensor fluctuations near the
horizon, expressed in the coordinates of an outside static
observer, in terms of a large correlations length. In this model
we deal with the mean value of these fluctuations which is taken
to be systematically determined by the outwards component of the
renormalized expectation value of the stress tensor of a quantum
field taken in some appropriately chosen quantum state. Therefore
the scaling law should basically understood as a condition imposed
on this state. In chapter 4 we present a dynamical derivation of
the scaling law on the basis of the backreaction effect using a
Planckian cutoff condition in the frame of an observer who uses
finite coordinates at the horizon. In chapter 5 we show that the
scaling law can be represented in form of a
fluctuation-dissipation theorem which relates the mean value of
the outwards fluctuations of the stress tensor near the horizon
to the black hole evaporation rate. Some arguments for deriving
corrections to the radiation temperature are then presented in
chapter 6. The paper ends with some concluding remarks.

\section{The Model}

We  consider
a two-dimensional analog of the Schwarzschild
black hole of mass $M$, described in coordinates which are
indicative of the outside observations (Schwarzschild coordinates)
by the metric
\begin{equation}
ds^{2}=-\Omega(r)dt^{2}+\Omega^{-1}(r)dr^{2},~~~\Omega(r)=1-2M/r,
\label{1}\end{equation} to which a massless scalar quantum field
$\phi$ is taken to be minimally coupled. Let the stress tensor of
$\phi$ in the (t,r) coordinates be denoted by $T_{\mu\nu}$. We
are primarily interested in the form of the low energy
fluctuations of $T_{\mu\nu}$ near the horizon, i.e. in the
effective horizon limit $r\rightarrow 2M$. In a semiclassical
treatment these fluctuations must be considered as random
fluctuations due to the horizon noise. Denoting their mean value
by $\delta T_{\mu\nu}$, the first fundamental task is how to
control the typical value of $\delta T_{\mu\nu}(r\rightarrow 2M)$
in terms of quantities accessible to a semiclassical treatment.

In a semicalssical theory the operator $T_{\mu\nu}$ arises as a
singular operator because it involves the product of the field
operator at a single point. Therefore one can generally assume
that the typical value of $\delta T_{\mu\nu}(r\rightarrow 2M)$
may be related to the effective horizon limit $r\rightarrow 2M$
of the renormalized expectation value
$<T_{\mu\nu}>^{ren.}_{\omega}$ taken in some appropriately chosen
quantum state $\omega$, namely
\begin{equation}
\delta T_{\mu\nu}(r\rightarrow 2M)\sim <T_{\mu\nu}(r\rightarrow
2M)>^{ren.}_{\omega}. \label{1-a}\end{equation} This relation
links two kinds of effects. The right hand side of (\ref{1-a}) is
the systematic value of the stress tensor near the horizon which
is accessible to a semiclassical treatment through the choice of
the quantum state $\omega$, whereas the left hand side is the
random value due to random fluctuations. The relation (\ref{1-a})
requires a systematic link between both values. The quantum state
$\omega$ is, therefore, assumed to link the random value of the
stress tensor $T_{\mu\nu}$ near the horizon with its systematic
counterpart, namely the renormalized expectation value of
$<T_{\mu\nu}(r\rightarrow 2M)>^{ren.}_{\omega}$. This has an
essential consequence for the characterization of the state
$\omega$.

\section{Long range scaling}

For the characterization of the state $\omega$ the determination
of the outwards component of $<T_{\mu\nu}>^{ren.}_{\omega}$ near
the horizon is very important because this component is the
indicative
quantity of the long-time observations in the outside
region. Let $<T_{uu}>^{ren.}_{\omega}$ denotes the outwards
component of $<T_{\mu\nu}>^{ren.}_{\omega}$ defined with respect
to the standard outward (retarded) time $u$ of the metric
(\ref{1}), namely
\begin{equation}
u=t-\stackrel{*}{r},~~~~\stackrel{*}{r}=r+2M\ln|\frac{r}{2M}-1|.
\label{2}\end{equation} The problem is how to determine the limit
$<T_{uu}(r\rightarrow 2M)>^{ren.}_{\omega}$. The relation
(\ref{1-a}) tells us that $<T_{uu}(r\rightarrow
2M)>^{ren.}_{\omega}$ is related to $\delta T_{uu}(r\rightarrow
2M)$ which is the mean value of the outwards fluctuations of the
stress tensor near the horizon. There is an argument, based on
the black hole causal structure, which suggests that these
fluctuations are correlated over all scales of lengths. The
argument goes as follows: The black hole causal structure implies
that null rays which are equispaced along the future null
infinity over long distances crowd up near the horizon over small
distances. This requires that the outwards fluctuations of the
stress tensor near the horizon shall be correlated over almost
all scales of lengths. In particular the typical scale of the
outward component of the renormalized stress tensor near the
horizon may be taken to be set by a large correlation length
$\xi$ which could have in principle its value many orders of
magnitude away from the Schwarzschid radius, giving us the the
scale hierarchy
\begin{equation}
\xi>>2M. \label{3}\end{equation}
The basic strategy is now to express
the limit $<T_{uu}(r\rightarrow 2M)>^{ren.}_{\omega}$ in terms
of the correlation length $\xi$ using dimensional arguments.
In doing this we should take into account
that, due to the scale hierarchy (\ref{3}),
any physical quantity may depend in principle
on the dimensionless ratio $2M/\xi$. For
the limit $<T_{uu}(r\rightarrow 2M)>^{ren.}_{\omega}$ we get therefore
on dimensional
grounds the general relation
\begin{equation}
<T_{uu}(r\rightarrow 2M)>^{ren.}_{\omega}=\xi^{-2}f(2M/\xi)
\label{z-4}\end{equation} where $f$ is a scaling function. The
particular choice of the scaling function $f$ depends on the
state $\omega$ and the corresponding Hilbert space. For
physically admissible states however we expect that the outwards
component of the renormalized stress tensor near the horizon does
not exhibit a significant sensitivity to a change  of the
dimensionless scaling variable $2M/\xi$ as long as this variable
remains small according to (\ref{3}). This means that the scaling
function may be approximated by a constant function. In this way
the geometric length $2M$ in the relation (\ref{z-4}) drops out
so that the significant scale is taken to be the correlation
length $\xi$ only. We are therefore led to predict the long range
scaling law
\begin{equation}
<T_{uu}(r\rightarrow 2M)>^{ren.}_{\omega}\sim \xi^{-2}.
\label{4}\end{equation} We should emphasize that the scaling law
(\ref{4}) as it stands is predicted on the basis of heuristic
arguments, and a systematic framework for its justification is
still missing. Regarding this point the following remark is
necessary. One may study the scaling law (\ref{4}) from the
viewpoint of the renormalization group arguments. For this
purpose the outwards fluctuations of the stress tensor expressed
in the coordinates of a static observer near the horizon should
properly be taken to be correlated over a static cutoff length
near the horizon, i.e. a cutoff length used by a static observer
properly located just outside the horizon. In this way the
correlation length $\xi$ appears to be systematically linked with
the value of such a static cutoff\footnote{Actually a static
cutoff length near the horizon is significantly large because a
static cutoff frequency for outgoing modes tends to zero at the
horizon [2]}. The requirement that an observable quantity such as
the left hand side of (\ref{4}) should be cutoff-independent
appears then to be in conflict with the scaling law (\ref{4}). In
the present case the inconsistency of (\ref{4}) with the standard
renormalization group arguments does not reflect a weakness of
our argumentations. In fact the standard renormalization group
arguments are not applicable in the present case where, due to the
black hole causal structure, a strong sensitivity of the outwards
fluctuations near the horizon to the cutoff mechanism can be
expected. This point is of particular importance for an
understanding of the scaling law (\ref{4}). It indicates that a
systematic framework for the justification of the scaling law
(\ref{4}) can not be based on the standard renormalization group
arguments, so another methods must be applied. We shall deal with
this issue in the next section where a dynamical derivation of
(\ref{4}) is given. This derivation is mainly based on two
assumptions, one involving a Planckian cutoff condition in the
frame of an observer who uses finite coordinates at the horizon
and one involving the backreaction effect. Although this
derivation may not seem to be conclusive, but it emphasizes the
fact that the scaling law (\ref{4}) may systematically be linked
with the physical mechanism of a cutoff. In the remaining part of
this section we collect some theoretical facts in connection with
(\ref{4}).

It is important to note that the scaling law (\ref{4}) implies
that the characteristic order of the magnitude of the expectation
value $<T_{uu}(r\rightarrow 2M)>^{ren.}_{\omega}$ is set by the
correlation length of the outwards fluctuations of the stress
tensor near the horizon, which is distinctly separated by the
scale hierarchy (\ref{3}) from the characteristic macroscopic
length of the system, namely $2M$. This feature implies that the
quantity $<T_{uu}(r\rightarrow 2M)>^{ren.}_{\omega}$ should
basically decouple from the dynamical constraint of the
renormalization theory describing the effective change of the
late-time configuration of the renormalized expectation value
$<T_{\mu\nu}>^{ren.}_{\omega}$, because this change can
generically be expected to occur on those typical macroscopic
length scales which are distinctly much smaller than $\xi$. That
this decoupling actually happens  to be the case may be seen from
the following
consideration:\\
The dynamical constraint of the renormalization theory can be
expressed in form of a hydrodynamic constraint, namely the
conservation law
\begin{equation}
\nabla^{\mu}<T_{\mu\nu}>^{ren.}_{\omega}=0.
\label{5}\end{equation}
This law can be used to determine the
static form of $<T_{\mu\nu}>^{ren.}_{\omega}$. In doing this we
ignore effects related to a preassigned time-dependence of the
expectation value $<T_{\mu\nu}>^{ren.}_{\omega}$. Naturally, we
assume that any time-dependence of $<T_{\mu\nu}>^{ren.}_{\omega}$
should be suppressed in the late-time limit. Therefore we look
for the static configuration of the expectation value
$<T_{\mu\nu}>^{ren.}_{\omega}$ which can be found to be [6][7]
\begin{equation}
<T_{\mu}^{\nu}>^{ren.}_{\omega}=T_{\mu}^{(1)\nu}+T_{\mu}^{(2)\nu}+T_{\mu}^{(3)\nu}
\label{6}\end{equation}
where in $(t,\stackrel{*}{r})$ coordinates
\begin{equation}
T_{\mu}^{(1)\nu}=
\left(\matrix{<T^{\alpha}_{\alpha}(r)>_{\omega}-\Omega^{-1}(r)H(r)&0\cr
0&\Omega^{-1}(r)H(r)~\cr}\right)
\label{6-a}\end{equation}
\begin{equation}
T_{\mu}^{(2)\nu}=
\Omega^{-1}(r)\frac{K}{M^{2}}\left(\matrix{1&-1\cr
1&-1\cr}\right)
\label{6b}\end{equation}
\begin{equation}
T_{\mu}^{(3)\nu}=
\Omega^{-1}(r)\frac{Q}{M^{2}}\left(\matrix{-2&0\cr
0&2\cr}\right).
\label{6c}\end{equation}
Here $K$ and $Q$ are arbitrary constants, and
\begin{equation}
H(r)=\frac{1}{2}\int^r_{2M}
\Big({{d\over{dr^\prime}}
\Omega(r^\prime)}\Big)<T^\alpha_\alpha(r^\prime)>_{\omega} dr^\prime.
\label{6d}\end{equation}
We can now determine the outwards component of this solution. We find
\begin{equation}
<T_{uu}>^{ren.}_{\omega}=\frac{1}{2}(H(r)+2Q/M^{2})-\frac{1}{4}\Omega(r)
<T^{\alpha}_{\alpha}(r)>_{\omega}
\label{7}\end{equation}
which near the horizon yields the scaling law
\begin{equation}
<T_{uu}(r\rightarrow 2M)>^{ren.}_{\omega}\rightarrow Q/M^{2}.
\label{8}\end{equation}
Thus, if the
conservation law is applied, it is possible to describe the
outwards component of $<T_{\mu\nu}>^{ren.}_{\omega}$ near the
horizon in terms of an integration constant which is not
dependent upon the particular dynamical coupling of $\phi$ to the
metric (\ref{1}). This implies that the specification of that
component must reflect a model independent general characteristic
of the vacuum state as observed by those observers which are
strictly localized  outside the black hole. This
observation establishes the decoupling of
$<T_{uu}(r\rightarrow 2M)>^{ren.}_{\omega}$ from
the dynamical constraint of the renormalization theory.

\section{The dynamical derivation of the scaling law}

The heuristic arguments that led to the scaling law (\ref{4}) can
find a dynamical justification by combining a cutoff condition
near the horizon with the backreaction effect of black hole
thermal radiation. To this aim we first note that the
unrenormalized expectation value $<T_{uu}(r\rightarrow
2M)>_{\omega}$ is mathematically a singular quantity. The
corresponding renormalized value can be obtained firstly by
introducing a Planckian cutoff length $l_c\sim 1$, and secondly by
specifying the reference frame to which the cutoff is applied.
The most natural reference frame for the imposition of a
Planckian cutoff is a reference frame of an infalling observer
crossing the horizon. The coordinate system which is indicative
of such an observer is a coordinate system which is finite at the
horizon, such as the inwards and the outwards Kruskal
time-coordinates, defined respectively as
\begin{equation}
V=4M~ exp(v/4M),~~~ U=-4M~ exp(-u/4M) \label{9}\end{equation}
where $v=t+\stackrel{*}{r}$ is the advanced time and $u$ is the
retarded time as given by (\ref{2}). In this coordinate system
the imposition of a Planckian cutoff on the expectation value
$<T_{uu}(r\rightarrow 2M)>_{\omega}$ is taken to correspond to
the requirement that the stress-tensor fluctuations near the
horizon shall be correlated over a cutoff length of Planckian
size. The correlation length of these fluctuations may therefore
be taken to be $l_c\sim 1$. This length sets the typical scale of
length for the determination of $<T_{UU}(r\rightarrow
2M)>^{ren.}_{\omega}$. Therefore on dimensional grounds we arrive
at the relation
\begin{equation}
<T_{UU}(r\rightarrow 2M)>^{ren.}_{\omega}\sim (l_c)^{-2} \sim 1
\label{10}\end{equation} which indicates that the state $\omega$
exhibits large stress-tensor fluctuations in the frame of an
infalling observer crossing the horizon. It is important to note
that this feature which arises from the cutoff condition reflects
the characteristic feature of the black hole complementarity
principle because it indicates that strong gravitational effects
may not decouple from the inside of the horizon, so that the
Hilbert space of the state $\omega$ becomes the wrong Hilbert
space for the low energy description of the physical state of an
infalling observer.

To derive the scaling law (\ref{4}) from the relation (\ref{10})
we first relate the
expectation value $<T_{UU}>^{ren.}_{\omega}$ to $<T_{uu}>^{ren.}_{\omega}$
using the coordinate transformation (\ref{9}) to find
\begin{equation}
<T_{UU}>^{ren.}_{\omega}=\frac{1}{4} ~exp(-r/M) V^{2} (r-2M)^{-2}
<T_{uu}>^{ren.}_{\omega}. \label{11}\end{equation}
We then try to
use this relation for the derivation of the scaling behavior of
$<T_{uu}>^{ren.}_{\omega}$ in the limit $r\rightarrow 2M$. The
first observation is that (\ref{11}) together with (\ref{10})
implies that $<T_{uu}>^{ren.}_{\omega}$ vanishes in the limit
$r\rightarrow 2M$. However this feature is an idealization of
neglecting the backreaction  due the Hawking effect. The
consideration of the backreaction implies that the effective
horizon limit $r\rightarrow 2M$ should be carried out with
respect to a mass scale which is slightly smaller than the mass
$M$, namely
\begin{equation}
r\rightarrow 2(M-\delta M)
\end{equation}
where $\delta M $ is of the order of the mass evaporated away
during the times just prior to the formation of the horizon. For
a sufficiently large black hole one can generally expect that
\begin{equation}
\delta M<< M
\end{equation}
holds. But it can be shown that $\delta M$ is even much smaller
than the Planck mass, namely $\delta M<<1$. To show this we
consider $M$ as a function of the advanced time $M(v)$, and let
$v_0$ be the value of the advanced time at which the horizon
would form if we neglect the backreaction effect. The mass
$\delta M$ can be estimated by
\begin{equation}
\delta M\sim |\frac{dM}{dv}(\tilde{v})| (v_0-\tilde{v})
\label{12}\end{equation}
where the time $\tilde{v}$ is taken to characterize the onset of the
evaporation process, so it must be very close to the horizon formation time
$v_0$. In general the time difference $\delta v=v_0-\tilde{v}$ must be
taken as much smaller than the characteristic time-scale of the system which is
set by the black hole mass. Therefore one should have
\begin{equation}
\delta v<< 2M.\label{13}\end{equation}
Although the validity of
this relation seems to be apparent from the context, but it can
be justified also by noting that, for a value $\tilde{v}$ of the
advanced  time characterized by (\ref{13}), a null-geodesic after
its propagation through a collapsing objects will be characterized
by a retarded time $\tilde{u}\sim -4M\ln{(v_0-\tilde{v})/2M}$;
which
is characteristic to the onset of the evaporation process [8][3].
The correct order of magnitude of $\delta M$ can now be
determined if we estimate $\frac{dM}{dv}(\tilde{v})$ by the
Hawking law
\begin{equation}
-\frac{dM}{dv}\sim \frac{1}{M^2}. \label{14}\end{equation} Using
this law in (\ref{12}) we obtain
\begin{equation}
\delta M\sim \frac{1}{M} \frac{\delta v}{M}
\label{15}\end{equation} which in conjunction with (\ref{13})
yields
\begin{equation}
\delta M<< 1. \label{16}\end{equation} This relation can be used
to estimate the renormalized expectation value
$<T_{uu}>^{ren.}_{\omega}$ near the horizon using the effective
horizon limit $r\rightarrow 2M-\delta M$ of the relation
(\ref{11}) together with (\ref{10}). We obtain
\begin{equation}
<T_{uu}(r\rightarrow 2M)>^{ren}_{\omega}\sim \xi^{-2}
\label{18}\end{equation}
where
\begin{equation}
\xi\sim\frac{2M}{\delta M}>>2M. \label{19}\end{equation} We
arrive therefore at the scaling law (\ref{4}).

\section{A fluctuation-dissipation theorem}

The scaling law (\ref{4})  links  the scaling behavior of the
renormalized expectation value $<T_{uu}>^{ren.}_{\omega}$ near
the horizon with the large correlations length $\xi$ of the
stress-tensor fluctuations near the horizon through the choice of
the quantum state $\omega$ in the outside region of the black
hole. One can therefore expect this state $\omega$ to posses a
dissipative character related to these fluctuations.
To see this we first write (\ref{4})
in the form
\begin{equation}
<T_{uu}(r\rightarrow 2M)>^{ren.}_{\omega}\sim\gamma M^{-2}
\label{corr}\end{equation} where $\gamma\sim(2M/\xi)^2<<1$. Now
combining (\ref{corr}) with the the Hawking law (\ref{14}) we get
\begin{equation}
<T_{uu}(r\rightarrow 2M)>^{ren.}_{\omega}\sim -\gamma
\frac{dM}{dv}. \label{20}\end{equation} This formulation of the
scaling law (\ref{4}) is instructive because it shows that the
quantum state $\omega$ links the outwards fluctuations of the
stress tensor near the horizon to a long-time dissipative
behavior, namely the evaporation rate of the black hole.
Therefore, the scaling law if combined with the Hawking law may
be brought into a form suggesting a fluctuation-dissipation
theorem. One can alternatively consider the arguments presented
in the previous chapter as demonstrating as to how such a theorem
can be derived on dynamical basis from a cutoff condition near
the horizon.

It is important to add the following remarks concerning the
correct physical interpretation of (\ref{20}). The fluctuations
that contribute to the left hand side of (\ref{20}) cause a small
disturbance of the black hole metric and the corresponding causal
structure. The relation (\ref{20}) describes how this disturbance
is related  to the evaporation rate via the choice of the quantum
state $\omega$ for the undisturbed system, i.e, the black hole
thermal state. This is is very much in the spirit of the general
framework of the response theory which relates the response of a
system to a small disturbance to the equilibrium characteristics
of the undisturbed system.

\section{Corrections to the Hawking effect}

It may be of interest to examine the effect of the scale-separation
$\xi>>2M$ on the Hawking effect. Generally, one expects to find
a deviation of the
black hole temperature from the Hawking temperature by a term
of the relative order $(2M/\xi)^\alpha$
where $\alpha$ is a characteristic exponent. To determine this exponent
we proceed as follows:
In two dimensions an outwards flux of thermal radiation
can be characterized at large $r$ by the energy momentum tensor
\begin{equation}
{\pi\over{12}}T^2\left(\matrix{-1&-1\cr
1&1\cr}\right)
\label{21}\end{equation}
in which $T$ is the temperature.
From (\ref{21}) one
infers
that a static, spherically symmetric
configuration of matter which describes the Hawking
radiation at large $r$ must have a stress tensor satisfying
the condition
\begin{equation}
T_{t}^{t}(r)\rightarrow T_{t}^{r}(r),~~~as ~r\rightarrow\infty
\label{22}\end{equation}
which means that the energy density and the flux are asymptotically equal.
If this condition is applied to the general solution (\ref{6}) one gets
\begin{equation}
K-\frac{1}{2}M^{2}[H(\infty)-<T_{\alpha}^{\alpha}(\infty)>_{\omega}+2Q]=0.
\label{23}\end{equation}

To derive the Hawking radiation from
such a relation one usually assumes two additional requirements.
The first one corresponds to the consistency of the trace anomaly
with respect to the two dimensional metric (\ref{1}), namely [6]
\begin{equation}
<T_{\alpha}^{\alpha}(r)>^{ren.}_{\omega}=\frac{M}{6\pi r^{3}}.
\label{24}\end{equation} The second requirement concerns the
finiteness of the energy momentum tensor at the horizon with
respect to a coordinate system which is finite there. In order to
implement the second assumption the standard derivation takes the
value $Q=0$ which arises as a pure effect of the transformation
law (\ref{11}) in the limit $r\rightarrow 2M$. There is however
some objections for considering the value $Q=0$ as the correct
one, because the finiteness condition of a quantum stress tensor
at the horizon requires us to investigate a cutoff condition in
the frame of an observer who uses finite coordinates at the
horizon, and we have seen that this leads to the scaling law
(\ref{4}) which together with (\ref{8}) predicts a non-vanishing
value $\sim (2M/\xi)^2$ for $Q$. Thus for the derivation of the
Hawking radiation we may take the consistency of the trace
anomaly together with this value of $Q$. At large $r$ the latter
condition predicts via the last term in (\ref{6}), namely the
tensor $T_{\mu}^{(3)\nu}$, a deviation of the thermal radiation
from the Hawking temperature of the relative order $(2M/\xi)^2$,
leading to the characteristic exponent $\alpha=2$ which coincides
in the present case with the dimensionality of space-time.

We should also remark that the non-vanishing value $Q\sim
(2M/\xi)^2$ predicts that the expectation value
$<T_{\mu\nu}>^{ren.}_{\omega}$ has at large $r$ a term
corresponding to a background heat bath with the temperature
$\sim 1/\xi$. This follows if one compares the tensor
$T_{\mu}^{(3)\nu}$ at large $r$ with the stress tensor of an
equilibrium gas, namely
\begin{equation}
{\pi\over{12}}(kT)^2\left(\matrix{-2&0\cr
0&2\cr}\right).
\label{15}\end{equation}
Such a model
may have some power at the cosmological level, in that
the background
heat bath may act as a  model for the thermal
equilibrium gas of an associated
cosmological horizon. In this way the cutoff condition on the horizon
may be linked with a small cosmological constant.

\section{Concluding remarks}

The paper has examined a new method for introducing a quantum
state $\omega$ for the outside region of a black hole. The
distinct feature of this method as compared with the standard
methods for introducing black hole states, such as those
discussed in [6] [7], is that it links via the scaling law
(\ref{4}) the choice of the quantum state $\omega$ outside the
horizon with the noise of gravitational effects inside
the horizon, and in this respect it emphasizes a general
relationship between a small disturbance of the black hole causal
structure and the choice of an external quantum state in the
absence of this disturbance. One may expect that the implications
that arise from this viewpoint may improve our uderstanding about
the nature of black hole evaporation.\\\\

{\bf Acknowledgement}\\
The author thanks the office of scientific research
of Sh.Beheshti University for financial support.\\\\


\begin{thebibliography}{99}

\bibitem{1} Fredenhagen K and Haag R, Commun. Math. Phys. 127, 273 (1990)
\bibitem{2} Jacobson Th,  Phys. Rev. D44, 1731 (1991)
\bibitem{3} Salehi H, Class. Quantum Grav. 10, 595 (1993)
\bibitem{4} Susskind L, Thorlacius L and Uglum J, Phys. Rev. D 48, 3743 (1993)
\bibitem{5} Lowe D, Polchinski J, Susskind L, Thorlacius L and Uglum J, Phys.Rev.
D 52, 6997 (1995)
\bibitem{6} Christensen S M, Fulling S A, Phys. Rev. D15, 2088 (1977)
\bibitem{7} Quantum fields in Curved Space, Birrell and Davies,
   Cambridge, England, (1982).
\bibitem{8} Hawking S, Commun. Math. Phys. 43, 199 (1975)

\end{thebibliography}
\end{document}